\documentclass[10pt,a4paper]{article}
\usepackage{jcappub}
\usepackage{amssymb,amsmath,amsfonts}
\usepackage{bm}
\usepackage{amsfonts}
\usepackage{latexsym}
\usepackage{graphicx}
\usepackage{amsmath}
\usepackage{palatino}
\usepackage{mathpazo}
\usepackage{textcomp}
\usepackage{subfigure}
\usepackage{float}
\usepackage{booktabs}
\usepackage{mathtools}
\usepackage{dcolumn}
\usepackage{ragged2e}

\usepackage{hyperref}

\arxivnumber{2605.22362}

\hypersetup{
    colorlinks=true,
    linkcolor=red,
    citecolor=blue,
    filecolor=magenta,      
    urlcolor=cyan,
    pdftitle={Constraining Spatial Curvature with Swampland-Motivated Priors},
    pdfpagemode=FullScreen} 
\usepackage{amsmath}
\usepackage{xcolor}
\usepackage{orcidlink}
\usepackage{tabularx}
\usepackage{array}

\title{Constraining Spatial Curvature with Priors from Swampland Conjectures}
\author[a]{Simran Arora\orcidlink{0000-0003-0326-8945}}
\author[b,a]{Hun Jang\orcidlink{0000-0002-9966-0775}}
\author[a,c,d]{Shinji Mukohyama\orcidlink{0000-0002-9934-2785}}
\affiliation[a]{Center for Gravitational Physics and Quantum Information, Yukawa Institute for Theoretical Physics, Kyoto University, Kyoto 606-8502, Japan}
\affiliation[b]{Center for Quantum Spacetime (CQUeST), Sogang University, Seoul 04107, Republic of Korea}
\affiliation[c]{Research Center for the Early Universe (RESCEU), Graduate School of Science, The University of Tokyo, Hongo 7-3-1, Bunkyo-ku, Tokyo 113-0033, Japan}
\affiliation[d]{Kavli Institute for the Physics and Mathematics of the Universe (WPI), The University of Tokyo Institutes for Advanced Study, The University of Tokyo, Kashiwa, Chiba 277-8583, Japan}

\emailAdd{arora.simran@yukawa.kyoto-u.ac.jp}
\emailAdd{hun.jang@nyu.edu}
\emailAdd{shinji.mukohyama@yukawa.kyoto-u.ac.jp}

\abstract{We study a string-motivated theoretical prior on the quintessential dark energy model with exponential potential, \( V(\phi) = V_0 e^{-\lambda \phi} \), allowing for non-zero spatial curvature. First, we formulate the corresponding dynamical system and investigate its cosmological evolution numerically, illustrating the phase-space behaviour and the influence of curvature on the background dynamics. In open universes (\( \Omega_k > 0 \)), it has been suggested that a curvature-related fixed point may support accelerated expansion even for relatively steep potentials compatible with swampland considerations. Next, we explicitly impose swampland-motivated priors on the slope parameter $\lambda$, restricting it to values consistent with the de Sitter conjecture that excludes the (curved) $\Lambda$CDM limit.  Furthermore, we restrict our considerations to the range of field excursion that is consistent with the swampland distance conjecture. Our primary interest is the possibility that such theoretically-motivated priors may shift values of cosmological parameters inferred by observational data, compared with the standard analysis based on theory-agnostic priors such as a sufficiently wide flat prior. We examine this possibility using a combination of Planck CMB data, DESI BAO measurements, and recent Type Ia supernova samples, performing a Bayesian inference of the model parameters. Our analysis indicates that the swampland-motivated prior mildly shifts the values of $\Omega_k$.}

\begin{document}

\begin{flushright} YITP-26-56, RESCEU-15/26, IPMU26-0022 \end{flushright}
\vspace{-8mm}

\maketitle

\section{Introduction}

A wide range of cosmological observations is accurately described by the $\Lambda$ Cold Dark Matter (CDM) model, in which late-time cosmic acceleration is attributed to a cosmological constant. Owing to its minimal parameterization and strong empirical success, this framework has long served as the standard model of precision cosmology. However, as contemporary surveys have reached a level of precision where systematic effects and internal consistency tests become increasingly important, the robustness of $\Lambda$CDM is being scrutinized through detailed cross-comparisons of independent high-precision probes. In this context, several persistent tensions have emerged and remained stable across successive data releases. These discrepancies suggest that, despite its overall success, the minimal late-time description may require a controlled extension, either through additional degrees of freedom or mild deviations from a strict cosmological constant, even if $\Lambda$CDM continues to provide a satisfactory global fit \cite{Verde:2019ivm,DiValentino:2021izs,Perivolaropoulos:2021jda,Abdalla:2022yfr,Aloni:2021eaq,Poulin:2023lkg,Vagnozzi:2023nrq,Kamionkowski:2022pkx}.

A prominent example is the Hubble-constant discrepancy between local distance-ladder determinations and the value inferred from cosmic-microwave-background (CMB) anisotropies within $\Lambda$CDM \cite{DiValentino:2021izs,Riess:2021jrx,Planck:2018vyg,Scolnic:2021amr,Brout_2022}. In addition, late-time structure probes in a number of analyses preferred a lower clustering amplitude (often summarized by $S_8$) than CMB-inferred expectations, with the reported significance depending on survey choice and modeling assumptions \cite{DES:2021Y3,KiDS:2020}. While these tensions do not uniquely identify the required modification, they single out late-time expansion history and structure growth as particularly sensitive directions in parameter space. Recent large-scale structure measurements have sharpened the empirical question of whether the dark-energy sector is strictly consistent with $\Lambda$ or instead exhibits mild time dependence. In particular, baryon acoustic oscillation (BAO) measurements from the Dark Energy Spectroscopic Instrument (DESI)~\cite{DESI:2024mwx,DESI:2025zgx} are well described by flat $\Lambda$CDM when considered independently. However, within extended dark-energy parameterizations, joint analyses with external datasets can shift relative to CMB-preferred constraints and indicate an apparent preference for dynamical dark energy.

Within the Chevallier--Polarski--Linder (CPL) parametrization, $w(a)=w_0 + w_a(1-a)$~\cite{Chevallier:2000qy,Linder:2002et,DES:2025upx,Hussain:2025nqy,Giare:2024gpk,Arora:2025msq,Chatrchyan:2024xjj,You:2025uon,Scherer:2025esj,Jiang:2024xnu}, DESI combinations with CMB and/or supernova (SNe) data can favor regions with $w_0 > -1$ and $w_a < 0$. Crucially, the statistical significance of these deviations from $\Lambda$CDM depends sensitively on the adopted SNe dataset. Analyses based on different supernova compilations, such as Pantheon+~\cite{Scolnic:2022PantheonPlus,Brout:2022PantheonPlusCosmo}, Union3~\cite{Rubin:2025Union3}, or the DES Year~5 sample (DES-SN5YR/DESY5) used in DESI analyses~\cite{DESI:2024mwx,DESI:2025zgx}, lead to noticeable shifts in the inferred cosmological parameters.

Here, a supernova compilation denotes an analysis-ready SNeIa Hubble-diagram dataset, defined not only by the selected sample but also by calibration procedures, light-curve standardization, bias corrections, and systematic covariance modeling. As a result, different compilations can induce systematic shifts in cosmological inference. Recent studies have emphasized that the apparent preference for evolving dark energy is entangled with supernova systematics, analysis priors, and dataset choices~\cite{Cortes:2024lgw,Efstathiou:2025MNRAS}, thereby limiting the robustness of current claims.

These considerations motivate treating dynamical dark-energy models, including effective field theory (EFT) descriptions, on an equal footing with $\Lambda$ in cosmological inference, while explicitly tracking dataset- and systematics-dependent effects~\cite{DESI:2025ExtendedDE,Escamilla:2023oce}. A closely related issue is parameter degeneracy between late-time dark-energy evolution and spatial curvature. At the level of observables, BAO and SNe constrain combinations of the Hubble rate and distance measures, notably the transverse comoving distance \cite{Hogg:1999ad}, whose derivation is shown in Appendix~\ref{appendix},
\begin{equation}
D_M(z)=\frac{c}{H_0}\,\frac{1}{\sqrt{|\Omega_k|}}\,
S_k\!\left(\sqrt{|\Omega_k|}\int_0^z\frac{dz'}{E(z')}\right),
\qquad
E(z)\equiv \frac{H(z)}{H_0},
\end{equation}
where $S_k(x)=\sin(x),\,x,\,\sinh(x)$ for $\Omega_k<0,\,=0,\,>0$, respectively. Since $D_M(z)$ and the luminosity distance $D_L(z)=(1+z)\,D_M(z)$ depend both on the line-of-sight integral of $1/E(z)$ and on curvature through the geometric mapping $S_k(\cdot)$, mild changes in the late-time expansion history (e.g.\ through $w(a)$) can partially compensate the curvature contribution. This is reflected in the strong degeneracy between dark-energy parameters and $\Omega_k$ in non-CMB curvature inferences based on late-time distance data \cite{Wu:2024faw}.

This entanglement is not merely formal: allowing $\Omega_k$ to vary can broaden or shift constraints on late-time parameters inferred from distance data, and curvature constraints become increasingly sensitive to assumptions about late-time dynamics when CMB information is weakened or replaced by non-CMB distance measurements \cite{Wu:2024faw}. In particular, analyses using non-CMB data combinations can yield curvature posteriors that shift significantly under extensions of the late-time model, illustrating directly the degeneracy between $\Omega_k$ and dark-energy phenomenology in distance-based inference. This motivates treating nonzero curvature as a minimal but physically meaningful extension of $\Lambda$CDM when assessing late-time deviations from $\Lambda$, since it captures a principal and observationally relevant degeneracy direction of BAO/SNe probes.

From a top-down perspective, string theory and related quantum-gravity considerations suggest that fully controlled (meta-)stable de Sitter vacua may be difficult to realize, motivating the swampland program and its cosmological implications \cite{Agmon:2022thq}. Among the conjectures most directly relevant to late-time acceleration are those constraining positive scalar potentials. The original de Sitter (dS) swampland conjecture posits a pointwise lower bound on the slope, 
\begin{equation}
%\frac{|\nabla V|}{V} \ \ge \ \frac{s_1}{M_{\rm P}} \,,
|\nabla V| \ \ge \ \frac{s_1}{M_{\rm P}}V \,,
\end{equation}
with an $\mathcal{O}(1)$ constant $s_1$, which would exclude de Sitter vacua if valid \cite{Obied:2018sgi}. A typical top-down motivation comes from dimensional reduction: under broad assumptions in higher-dimensional (super) gravity compactifications (e.g.\ energy-condition inputs), constraints on accelerated expansion in the lower-dimensional Einstein frame can be translated into lower bounds on gradients of the effective potential along moduli directions \cite{Obied:2018sgi}~\footnote{See \cite{Mizuno:2019pcm} for a generalization of the de Sitter conjecture to a field with a nonlinear kinetic Lagrangian such as a DBI scalar.}. In this interpretation, $s_1$ parametrizes the strength of a UV-motivated consistency requirement; if late-time acceleration is realized by rolling scalar dynamics rather than a strict cosmological constant, such potential-based criteria act directly on the dark-energy EFT.

Pointwise slope conditions could be missing something and inconsistent with what we see in the nature such as the Higgs potential~\cite{Denef:2018etk,Murayama:2018lie}, motivating refined statements incorporating second-derivative information. Refined or combined de Sitter swampland criteria allow (depending on the precise formulation) either a sufficiently large gradient or the presence of a sufficiently tachyonic direction in the Hessian, encoded for example through bounds involving the minimal eigenvalue of $\nabla_i\nabla_j V$ \cite{Andriot:2018mav,Andriot:2018wzk,Garg:2018reu}. Given ongoing debate regarding the quantitative status and the order-one parameters appearing in these conjectures, we treat them as conditional theoretical inputs rather than empirical statements.

A complementary set of UV-motivated constraints arises from towers of light states and the Trans-Planckian Censorship Conjecture (TCC). In asymptotic regions of moduli space, towers of states whose characteristic mass scale decreases exponentially with geodesic field distance are expected in many string-theoretic settings, potentially limiting the regime of validity of a single-field EFT \cite{Agmon:2022thq}. The TCC can be formulated as the requirement that trans-Planckian modes do not exit the Hubble horizon \cite{Bedroya:2019snp}; applied to scalar-driven expansion in asymptotic regimes, it yields constraints on the asymptotic behaviour of the potential. Quantum corrections can also affect bounds obtained from classical arguments, motivating scenario-dependent candidates for $s_1$ \cite{Sun:2019obt}. Recent work has further emphasized that the dynamical impact of light towers in FLRW cosmologies can constrain the viability of certain attractor solutions from the EFT perspective \cite{Casas:2024oak}. Related top-down settings, such as string-gas constructions, provide additional parametric expectations and explicit realizations consistent with dS-type criteria \cite{Laliberte:2019sqc}.

These considerations suggest a clear statistical role for swampland conjectures in cosmological inference. Since the conjectures represent proposed UV-consistency requirements rather than observational information, we implement them as ``theory priors'' acting on the parameter space, rather than as targets to be confronted with the posterior distributions. In Bayesian terms, with cosmological and nuisance parameters $\theta$ and datasets $D$, a swampland criterion as the theory prior can be encoded as a factor $\pi_{\rm sw}(\theta)$ that restricts or re-weights an underlying prior $\pi(\theta)$, so that the corresponding posterior $P(\theta \,|\, D,{\rm sw})$ is given by
\begin{equation}
P(\theta \,|\, D,{\rm sw}) \ \propto\ \mathcal{L}(D\,|\,\theta)\, \pi(\theta)\, \pi_{\rm sw}(\theta)\,,
\end{equation}
where $\pi_{\rm sw}$ summarizes UV-motivated consistency information logically independent of the statistical model for the data \cite{Agmon:2022thq}. Because the order-one parameters entering these conjectures are not uniquely fixed, we do not adopt a single numerical choice; instead we consider a representative set of theoretically motivated candidates corresponding to different assumptions and scenarios, including compactification-based estimates \cite{Obied:2018sgi}, quantum-corrected variants \cite{Sun:2019obt}, TCC-driven asymptotic bounds \cite{Agmon:2022thq}, light-tower cosmology considerations \cite{Casas:2024oak}, and string-gas-inspired expectations \cite{Laliberte:2019sqc}. 

With these theory priors specified, we confront the resulting model space with cosmological observations and address the following question: within the quintessential dark energy model extended to allow nonzero spatial curvature, {\it which curvature branch is most compatible with swampland-motivated UV-consistency requirements while remaining consistent with the data?} Here ``branch'' refers to the sign of $\Omega_k$ (open, $\Omega_k>0$, versus closed, $\Omega_k<0$) and, more generally, to the possibility of multi-modal posteriors in curvature-extended fits. Since curvature is observationally entangled with mild departures from $\Lambda$ in late-time distance measurements, this provides a controlled setting in which to quantify how UV-motivated consistency conditions can influence the interpretation of emerging late-time trends. For recent reviews on dark energy in the context of string theory, including discussions of exponential quintessence models, see Refs.~\cite{Wu:2026qog,Han:2018yrk,Bhattacharya:2024hep,Gialamas:2024lyw,Andriot:2026lac,Andriot:2024sif,Dinda:2025iaq}.

The paper is organized as follows. In Sec.~\ref{conjectures}, we review the relevant swampland conjectures and discuss their implications for the theoretical priors. In Sec.~\ref{sec3}, we present the dynamical system analysis for the curved quintessence model, followed by a discussion of its consistency with the swampland conjectures in Sec.~\ref{sec4}. In Sec.~\ref{Data}, we describe the cosmological datasets used in our analysis and present the resulting observational constraints. Finally, we summarize our main findings in Sec.~\ref{conc}.

\section{Swampland conjecture as prior} \label{conjectures}

From a top-down perspective, we view the swampland conjectures as {\it UV consistency conditions} constraining low-energy scalar-field effective field theories coupled to gravity. Accordingly, in our cosmological inference they enter as {\it theory priors} that restrict (or reweight) the {\it a priori} allowed parameter space, and the role of cosmological datasets is to update these UV-motivated priors into posteriors. Concretely, we implement the swampland constraints as bounds on the shape of positive scalar potentials and on their asymptotic behavior, and then ask how these consistency requirements reshape the inferred cosmological parameter space. In particular, which sign and magnitude of the spatial curvature is most compatible with the data once the UV-consistency priors are imposed.

In this section, we therefore (i) summarize the specific conjectures used in this work, namely the (refined/combined) de Sitter criteria and the asymptotic trans-Planckian censorship bounds, (ii) introduce the corresponding order-one parameters ($s_1$, $s_2$, and related combinations) that quantify the strength of the constraints, and (iii) compile a set of theoretically motivated candidate values for $s_1$ arising from distinct scenarios (Table~\ref{tab:s1_scenarios}). Since these parameters are not uniquely fixed and their precise values remain debated, we do not adopt a single choice; instead, we explore a representative value of candidates as alternative theory priors and examine the robustness of the resulting cosmological preference for spatial curvature across them.

The precise numerical implementation of de Sitter-type swampland bounds has been discussed in several closely related forms in the literature. In the original de Sitter conjecture, the condition $|\nabla V| \geq cV$ was proposed with $c$ taken to be a positive coefficient of order unity, without specifying a unique numerical value \cite{Obied:2018sgi}. Subsequent works developed refined versions of this criterion, incorporating both the first and second derivatives of the scalar potential \cite{Ooguri:2018wrx}, as well as alternative refinements involving additional parameters \cite{Andriot:2018wzk}. TCC-motivated analyses and studies in different spacetime dimensions have further investigated how these coefficients may be constrained or related under various assumptions \cite{Andriot:2020lea,Andriot:2022xjh}. Consequently, while the strong/asymptotic de Sitter Swampland conjecture suggests the natural reference value $s_1=\sqrt{2}$ in four dimensions, our statement was intended to refer more broadly to the range of de Sitter-type swampland constraints explored in the literature, rather than to question this particular value.

To make this strategy concrete, we parametrize the strength of the UV-consistency requirement by the order-one coefficients appearing in the conjectural bounds, most notably the slope parameter $s_1$ (and, when relevant, $s_2$ and combinations thereof). Since these coefficients are not uniquely fixed and depend on the underlying derivational assumptions, we adopt a scenario-based approach: we compile representative theoretical estimates for $s_1$ arising in different top-down settings and use them as alternative theory priors. Table~\ref{tab:s1_scenarios} summarizes the scenarios considered in this work together with their corresponding predictions for the lower bound parameter $s_1$.

\begin{table}[t]
\centering
\resizebox{\textwidth}{!}{%
\begin{tabular}{@{} p{5cm} c p{5cm} r @{}}
\toprule
\textbf{Scenario} &
\textbf{Theoretical value of $s_1$} &
\textbf{Notes} &
\textbf{Reference} \\
\midrule
({\bf S1}) Compactification of $D$-dimensional supergravity to $d$ dimensions (with strong energy condition) &
$ s_1 = 2\sqrt{\dfrac{(D-2)}{(D-d)(d-2)}} $ &
--- &  Obied {\it et al.} \cite{Obied:2018sgi} \\[8pt]

({\bf S2}) Compactification of $D$-dimensional supergravity to $d$ dimensions (with null energy condition) &
$ s_1 = 2\sqrt{\dfrac{(D-d)}{(D-2)(d-2)}} $ &
--- & Obied {\it et al.} \cite{Obied:2018sgi} \\[8pt]

({\bf S3}) Quantum correction to strong energy condition ($D \to 4$) &
$ s_1 = \sqrt{\dfrac{2(D-2)}{(D-4)}}\!\left(1+\dfrac{\beta}{3\pi}\right) $ &
$\beta \sim \mathcal{O}(1)$ from $S_{\text{out}}(R)\!\sim\!\alpha + \beta R^2H^2$ &
Sun {\it et al.} \cite{Sun:2019obt} \\[8pt]

({\bf S4}) Quantum correction to null energy condition ($D \to 4$) &
$ s_1 = \sqrt{\dfrac{2(D-4)}{(D-2)}}\!\left(1+\dfrac{C\beta}{3\pi}\right) $ &
$C=\!\!\int d^{D-4}\!y\,\sqrt{h}\,\Omega^{-2\gamma}$ &
Sun {\it et al.} \cite{Sun:2019obt} \\[8pt]

({\bf S5}) TCC (tower lighter than Hubble) &
$ s_1 = \dfrac{2}{\sqrt{(d-1)(d-2)}} $ &
--- & Agmon {\it et al.} \cite{Agmon:2022thq} \\[8pt]

({\bf S6}) TCC (tower heavier than Hubble) &
$ s_1 = \dfrac{2}{\sqrt{d-2}} $ &
--- & Agmon {\it et al.} \cite{Agmon:2022thq} \\[8pt]

({\bf S7}) Sharpened distance and dS conjectures &
$ s_1 = \dfrac{2}{\alpha(d-2)}, \quad \alpha \geq \tfrac{1}{\sqrt{d-2}} $ &
$\alpha$ controls slope bound &
Casas and Ruiz \cite{Casas:2024oak} \\[8pt]

({\bf S8}) String gas setting &
$ s_1 = \dfrac{1}{\sqrt{D-d}} $ &
$D-d$ compactified dimensions &
Laliberte and Brandenberger \cite{Laliberte:2019sqc} \\[8pt]

({\bf S9}) Combined dSC &
$ s_1^c + p_2(1+s_2) \geq 1 $ &
%$p_2 = \dfrac{c\lambda^{c-1}}{(1+3\lambda)}$ 
--- &
Andriot and Roupec \cite{Andriot:2018mav} \\
\bottomrule
\end{tabular}%
}
\caption{Summary of compactification and conjecture scenarios with corresponding theoretical values of $s_1$.}
\label{tab:s1_scenarios}
\end{table}

For later convenience, we also evaluate these expressions for the representative choices $(D,d)=(10,4)$ and $(11,4)$, corresponding to common string/M-theory inspired higher-dimensional origins with a four-dimensional effective description. The results are shown in Table~\ref{tab:s1_scenarios_eval_cols}, which makes explicit the relative strength of the resulting bounds across scenarios.

\begin{table}[t]
\centering
\begin{tabularx}{\textwidth}{@{} 
    >{\raggedright\arraybackslash}X 
    >{\centering\arraybackslash}m{3.5cm} 
    >{\centering\arraybackslash}m{3.5cm} 
@{}}
\toprule
\textbf{Scenario} &
\textbf{$D{=}10,\ d{=}4$} &
\textbf{$D{=}11,\ d{=}4$} \\
\midrule

({\bf S1}) Compactification of $D$-dimensional supergravity to $d$ dimensions (strong energy) &
$ s_1 = 2\sqrt{\tfrac{2}{3}} $ &
$ s_1 = 2\sqrt{\tfrac{9}{14}} $ \\[5pt]

({\bf S2}) Compactification of $D$-dimensional supergravity to $d$ dimensions (null energy) &
$ s_1 = \sqrt{\tfrac{3}{2}} $ &
$ s_1 = 2\sqrt{\tfrac{7}{18}} $ \\[5pt]

({\bf S3}) Quantum correction to strong energy condition ($D \to 4$) &
$ s_1 = \sqrt{\tfrac{8}{3}}\!\left(1+\tfrac{\beta}{3\pi}\right) $ &
$ s_1 = \sqrt{\tfrac{18}{7}}\!\left(1+\tfrac{\beta}{3\pi}\right) $ \\[5pt]

({\bf S4}) Quantum correction to null energy condition ($D \to 4$) &
$ s_1 = \sqrt{\tfrac{3}{2}}\!\left(1+\tfrac{C\beta}{3\pi}\right) $ &
$ s_1 = \sqrt{\tfrac{14}{9}}\!\left(1+\tfrac{C\beta}{3\pi}\right) $ \\[5pt]

({\bf S5}) TCC (tower lighter than Hubble) &
$ s_1 = \sqrt{\tfrac{2}{3}} $ &
$ s_1 = \sqrt{\tfrac{2}{3}} $ \\[6pt]

({\bf S6}) TCC (tower heavier than Hubble) &
$ s_1 = \sqrt{2} $ &
$ s_1 = \sqrt{2} $ \\[6pt]

({\bf S7}) Sharpened distance and strong de Sitter conjectures &
$ s_1 = \tfrac{1}{\alpha},\ \alpha \ge \tfrac{1}{\sqrt{2}} $ &
$ s_1 = \tfrac{1}{\alpha},\ \alpha \ge \tfrac{1}{\sqrt{2}} $ \\[5pt]

({\bf S8}) String gas setting &
$ s_1 = \tfrac{1}{\sqrt{6}} $ &
$ s_1 = \tfrac{1}{\sqrt{7}} $ \\[5pt]

({\bf S9}) Combined dSC &
$ s_1^{\,c} + p_2(1+s_2) \geq 1 $ &
$ s_1^{\,c} + p_2(1+s_2) \geq 1 $ \\

\bottomrule
\end{tabularx}

\caption{Values of $s_1$ for each scenario evaluated at $(D,d)=(10,4)$ and $(11,4)$. The ordering of magnitudes is $\mathbf{S8}<\mathbf{S5}<\mathbf{S2}<\mathbf{S6}<\mathbf{S1}$, with $\mathbf{S7} \leq \sqrt{2}$ for both $D=10$ and $D=11$. We plan to adopt $\mathbf{S2}$ for our analysis in Sec. 5.} 
\label{tab:s1_scenarios_eval_cols}
\end{table}

Having specified the set of theoretically motivated candidates for $s_1$, we next state the conjectural inequalities that we impose as priors. Throughout, we emphasize that these conditions are treated as UV-consistency criteria that restrict the admissible EFT parameter space, while the observational likelihood subsequently determines which regions of this restricted space are preferred.

To do this, we focus on the following swampland conjectures. The {\bf de Sitter (dS) conjecture} is that a scalar potential has to satisfy 
\begin{eqnarray}
%\frac{|\nabla V|}{V} \geq \frac{s_1}{M_{\rm P}}
|\nabla V| \geq \frac{s_1}{M_{\rm P}}V\,, \label{dSconjecture}
\end{eqnarray}
where $|\nabla V| \equiv \sqrt{g^{ij}\partial_{\phi_i}V\partial_{\phi_j}V}$.
This bound is %intended to apply in regions of field space with $V>0$, 
thereby constraining the possibility of realizing sustained acceleration from a positive scalar potential. In our analysis, it functions as a prior restriction on the model space: for a given scenario (i.e.\ a chosen $s_1$), parameter combinations that imply an effective potential violating the inequality are excluded (or strongly disfavored) at the prior level.

Since a small gradient can occur near critical points of the potential without implying a controlled (meta-)stable de Sitter vacuum, refined formulations incorporate second-derivative information by constraining the most tachyonic direction of the Hessian. In particular, one often considers the {\bf refined dS conjecture},
which states that the potential should satisfy either (\ref{dSconjecture}) or
\begin{eqnarray}
\textrm{min}(\nabla_i\nabla_j V) \leq -\frac{s_2}{M_{\rm P}^2}V.  \label{refined-dSconjecture}   
\end{eqnarray}
Here $s_2$ is another order-one positive coefficient, and $\textrm{min}(\nabla_i\nabla_j V)$ denotes the minimum eigenvalue of the covariant Hessian in field space. As reviewed in \cite{Lehnert:2025izp}, refined criteria involve subtleties both in their regime of validity and in their precise parametrization; accordingly, in the present work we do not attempt to infer $(s_1,s_2)$ from data, but instead treat them as external theoretical inputs defining alternative priors. For a positive exponential potential, (\ref{refined-dSconjecture}) is never satisfied, and thus the refined dS conjecture is reduced to the original dS conjecture (\ref{dSconjecture}).

To mitigate the limitations of using either the pure-slope or pure-Hessian condition in isolation, we could also consider a combined criterion that interpolates between the two types of information in a single inequality. The {\bf combined dS conjecture} \cite{Andriot:2018mav} is given by
\begin{eqnarray}
\Big(M_{\rm P}\frac{|\nabla V|}{V}\Big)^c -p_2 M_{\rm P}^2 \frac{\textrm{min}(\nabla_i\nabla_j V)}{V} \geq p_1\,, \quad \mbox{or}\quad V\leq 0\,,
\end{eqnarray}
where %$c>2$, $p_1,p_2 >0$, and $p_1+p_2=1$ which should be satisfied at any point in field space where the potential is positive. In this expression, 
$c>2$ and $(p_1,p_2)$ are positive weights satisfying $p_1+p_2=1$. The first inequality is designed to %be satisfied whenever $V>0$, and 
effectively enforces that either the normalized slope is large enough or the potential exhibits a sufficiently tachyonic direction, in a manner quantified by the chosen parameters \cite{Andriot:2018mav}. In terms of the parameters $s_1$ and $s_2$, the combined dS conjecture can be rewritten as \cite{Ahmed:2024rdd}
\begin{eqnarray}
    s_1^c + p_2(1+s_2) \geq 1\,,
\end{eqnarray}
where $0<p_2<1$.
This rewriting makes transparent how the combined criterion can be mapped onto an effective constraint on the slope parameter once the relevant class of potentials is specified. In our scenario-based implementation, we use this relation to motivate an additional family of priors (scenario S9 in Table~\ref{tab:s1_scenarios}) that encodes the combined dS condition in terms of $s_1$ and the model-dependent weight $p_2$ \cite{Andriot:2018mav}.
%where $p_2 = \dfrac{c\lambda^{c-1}}{(k+1+2\lambda)}$ with respect to the scalar potential $V(\phi)=(n(\phi) e^{d\phi})^{-k}$ where $\lambda=kd$. When we set $d=1$, we have $\lambda=k$, so that $p_2 = \dfrac{c\lambda^{c-1}}{(1+3\lambda)}$. What the values of $s_1$ and $s_2$ should be is an ongoing debate. However, we give an overview of the possible candidates for the values.
%Given this theoretical uncertainty, our strategy is to treat the set of values in Table~\ref{tab:s1_scenarios} as representative UV-motivated priors and to test which regions of cosmological parameter space remain viable and are preferred by the data under each choice, rather than committing to a single universal value.
For a positive exponential potential, the combined dS conjecture is reduced to the original dS conjecture with a different choice of $s_1$.

In asymptotic regions of field space (the infinite-distance limit), towers of states are expected to become light and may affect the cosmological evolution, potentially challenging the validity of a single-field EFT description \cite{Agmon:2022thq}. In this setting, the {\bf trans-Planckian censorship conjecture (TCC)} can be translated into an asymptotic lower bound on the normalized slope of the potential~\footnote{For other cosmological implications of the TCC, see \cite{Mizuno:2019bxy} and references therein.}. When additional degrees of freedom associated with the light tower contribute to the background expansion, one obtains the milder asymptotic bound
\begin{eqnarray}
    \frac{|\nabla V|}{V} \bigg|_{\rm asympt} \geq \frac{2}{\sqrt{(d-1)(d-2)}}.
\end{eqnarray}
where the subscript ``$\rm{asympt}$'' means the infinite distance limit, $\Delta\phi \rightarrow \infty$.
This case corresponds to scenario S5 in Table~\ref{tab:s1_scenarios}.
If instead the evolution is effectively single-field, in the sense that the relevant tower does not contribute appreciably to the background expansion, the TCC implication for the asymptotic slope becomes stronger, yielding \cite{Agmon:2022thq}
\begin{eqnarray}
       \frac{|\nabla V|}{V} \bigg|_{\rm asympt} \geq \frac{2}{\sqrt{d-2}}. 
\end{eqnarray}
This bound is also known as the Strong de Sitter conjecture, originally proposed in Refs.~\cite{Rudelius:2021oaz,Rudelius:2021azq}, where it was formulated for general scalar potentials. In the asymptotic limit, it coincides with scenario S6 in Table~\ref{tab:s1_scenarios}.

We thus obtain a set of explicit, scenario-dependent inequalities that encode UV-consistency requirements on the effective dark-energy sector. In the remainder of the paper, we implement one of these conditions as a theory prior in our parameter inference and quantify how late-time cosmological data update it into posterior. The key diagnostic is how such UV-motivated prior reshapes the allowed region of parameter space-most notably the inferred spatial curvature-thereby enabling a concrete comparison between swampland-motivated consistency conditions and cosmological observations.
~~\\
\noindent {\bf Scenario ``S2" as the reference benchmark.} We take the scenario {\bf S2} in Table \ref{tab:s1_scenarios_eval_cols} as the reference benchmark, to be employed as a {\it theory prior} in the cosmological analysis done in Sec.~\ref{Data}. The rationale is that {\bf S2} is based on the null energy condition, whereas {\bf S1} and {\bf S3} depend on a more restrictive energy-condition assumption in the compactification framework of interest. Accordingly, {\bf S2} yields a comparatively conservative classical bound with minimal additional structure. This makes it a natural baseline for comparison: {\bf S4} represents the corresponding quantum-corrected NEC-based scenario, {\bf S5} and {\bf S6} incorporate further assumptions about the hierarchy between the tower mass and Hubble scales, and {\bf S7}--{\bf S9} invoke more specialized ingredients such as sharpened swampland constraints, string-gas dynamics, or combined de Sitter conditions. We therefore regard {\bf S2} as a sufficiently simple reference point from which the impact of increasingly restrictive or model-dependent assumptions can be evaluated.

\section{Cosmological analysis} \label{sec3}
 
We consider the system corresponding to a universe driven by a single canonical quintessence scalar field acting as dark energy \cite{Ratra:1987rm}. The background dynamics are studied within a four-dimensional cosmological framework that includes radiation, pressureless matter, spatial curvature, and the scalar field component. The action for a minimally coupled canonical scalar field in the presence of non-relativistic matter is given by
\begin{eqnarray}
    S = \int d^4x \sqrt{-g} \left[ \frac{1}{2}M_{p}^2\,R - \frac{1}{2}g^{\mu \nu}\partial_{\mu}\phi\partial_{\nu}\phi - V(\phi) + \mathcal{L}_m \right],
\end{eqnarray}
where $M_p = 1/\sqrt{8\pi G_{N}}$ denotes the reduced Planck mass. 
We assume a homogeneous and isotropic Friedmann–Lemaître–Robertson–Walker (FLRW) spacetime with the line element
\begin{equation}
    ds^2 = -dt^2 + a(t)^2 \left( \frac{dr^2}{1-k r^2} + r^2 d\Omega^2  \right),
\end{equation}
where $a(t)$ denotes the scale factor and $k = 0, \pm 1$ characterizes the three-dimensional spatial curvature. The matter sector consists of a canonically normalized scalar field $\phi$ with scalar potential $V(\phi)$, together with perfect-fluid components representing radiation and non-relativistic matter. Each fluid satisfies an equation of state of the form $p_{i}=w_{i}\rho_{i}$, where $i=r,m$ corresponds to radiation and matter. In this case, the equations of motion can be written as follows: 
\begin{eqnarray}
 &&   H^2 = \frac{1}{3} \sum_i \rho_{i} \ , \quad \quad  \text{where $i=r,m,k,\phi$} \\
 &&   \dot{H} = -\frac{1}{2} \sum_i(\rho_{i}+ p_{i}) \ ,\\
 &&   \ddot{\phi} + 3 H \dot{\phi} + V_{,\phi} = 0 \ .
\end{eqnarray}
Here, we set the reduced Planck mass $M_{p}=1$ and an overdot denotes differentiation with respect to cosmic time with $H=\frac{\dot{a}}{a}$. For each cosmological component, we define the corresponding density parameter as  
\begin{equation}
\Omega_i \equiv \frac{\rho_i}{3H^2} \ .
\end{equation}
In terms of these variables, the first Friedmann equation can be written as
\begin{equation}
1 = \sum_i \Omega_i \ ,
\label{friedmann_density_sum}
\end{equation}
which may equivalently be expressed as
\begin{equation}
1 - \Omega_k = \Omega_m + \Omega_r + \Omega_{\phi} = \Omega_{T} \ , \quad \text{where} \quad \Omega_k \equiv -\frac{k}{a^2 H^2} \ .
\end{equation}

Equation~\eqref{friedmann_density_sum} makes explicit the geometrical role of spatial curvature. An open universe (\( k = -1 \)) corresponds to \( \Omega_k > 0 \) and therefore \( \Omega_T < 1 \), while a closed universe (\( k = +1 \)) implies \( \Omega_k < 0 \) and hence \( \Omega_T > 1 \). The spatially flat case (\( k = 0 \)) is recovered when \( \Omega_k = 0 \), for which \( \Omega_T = 1 \).

Finally, we can define the effective equation of state of the full system as 
\begin{eqnarray}
    w_{eff} = \frac{p_{eff}}{\rho_{eff}} = w_{\phi}\,\Omega_{\phi} - \frac{1}{3}\Omega_{k} + \frac{1}{3}\Omega_{r} \ .
\end{eqnarray}

We now examine the cosmological evolution within the dynamical systems framework. Rather than performing a stability analysis, which has been carried out in literature \cite{Andriot:2024jsh,Gosenca:2015qha,vandenHoogen:1999qq,Andriot:2023wvg,Alestas:2024gxe}, we focus on the numerical integration of the autonomous system in order to directly track the background evolution across different cosmological epochs.

\begin{eqnarray}
    x = \frac{\dot{\phi}}{\sqrt{6}H} \ , \quad y = \frac{\sqrt{V}}{\sqrt{3}H} \ , \quad z = \frac{\sqrt{-k}}{a H} \ , \quad u = \frac{\sqrt{\rho_r}}{\sqrt{3}H} \ . \label{dynamic_sys_variables}
\end{eqnarray}
In the following, we restrict our analysis to non-positively curved geometries, \( k \leq 0 \), corresponding to spatially flat \( k = 0 \) and open \( k = -1 \) cases. We further assume a non-negative scalar potential \( V(\phi) \geq 0 \), ensuring a physically well-defined energy density.
The cosmological dynamics can be reformulated as an autonomous system by combining the second Friedmann equation with the scalar-field equation of motion. For concreteness, we consider an exponential potential of the form 
\begin{equation}
V(\phi) = V_0 \, e^{-\lambda \phi} ,
\end{equation}
where \( \lambda > 0 \) characterizes the slope of the potential and \( V_0 \ge 0 \). Exponential potentials and their scaling solutions have been extensively studied in Refs.~\cite{Copeland:1997et,Ferreira:1997hj,Liddle:1998xm,Capozziello:2005ra,Akrami:2025zlb}. Introducing suitable dimensionless variables, the background evolution equations can be cast into the following autonomous system:
\begin{align}
\label{d1}
x' &= \sqrt{\frac{3}{2}}\, y^2 \lambda  
+ x \left[ 3(x^2 - 1) + z^2 + \frac{3}{2}\Omega_m + 2u^2 \right], \\ \label{d2}
y' &= y \left[ -\sqrt{\frac{3}{2}}\, x \lambda 
+ 3x^2 + z^2 + \frac{3}{2}\Omega_m + 2u^2 \right], \\ \label{d3}
z' &= z \left[ z^2 - 1 + 3x^2 + \frac{3}{2}\Omega_m + 2u^2 \right], \\ \label{d4}
u' &= u \left[ z^2 - 2 + 3x^2 + \frac{3}{2}\Omega_m + 2u^2 \right].
\end{align}

In terms of these variables, the scalar-field density parameter and equation-of-state parameter are given by
\begin{equation}
\Omega_\phi = x^2 + y^2 ,
\qquad
w_\phi = \frac{x^2 - y^2}{x^2 + y^2} ,
\qquad
\Omega_k = z^2 ,
\qquad
\Omega_r = u^2 ,
\qquad 
w_{eff} = x^2 - y^2 - \frac{z^2}{3} +\frac{u^2}{3}.
\label{omegas_def}
\end{equation}
The first Friedmann equation imposes the constraint
\begin{equation}
\Omega_m = 1 - x^2 - y^2 - z^2 - u^2 \,,
\label{Omega_m_constraint}
\end{equation}
which closes the system by expressing the matter density parameter in terms of the dynamical variables. As a result, Eq~\eqref{Omega_m_constraint} defines an invariant hypersurface in phase space, effectively reducing the dimensionality of the autonomous system by one.

\begin{figure*}[h!]
\centering
\includegraphics[scale = 0.36]{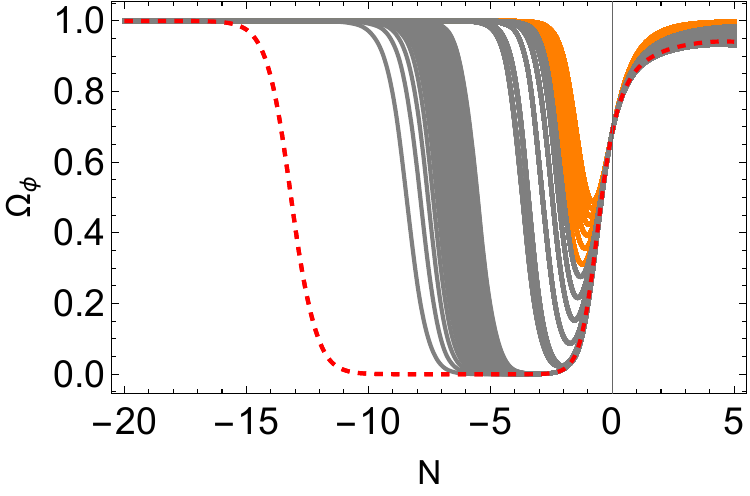} \hspace{0.1in}
\includegraphics[scale = 0.38]{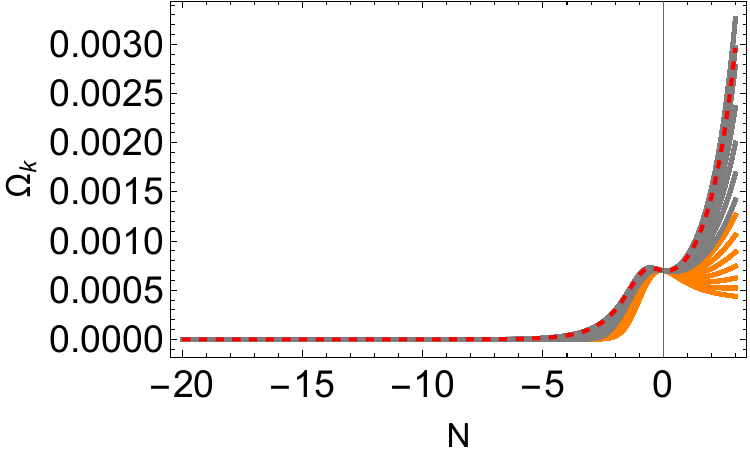}  \hspace{0.1in}
\includegraphics[scale = 0.36]{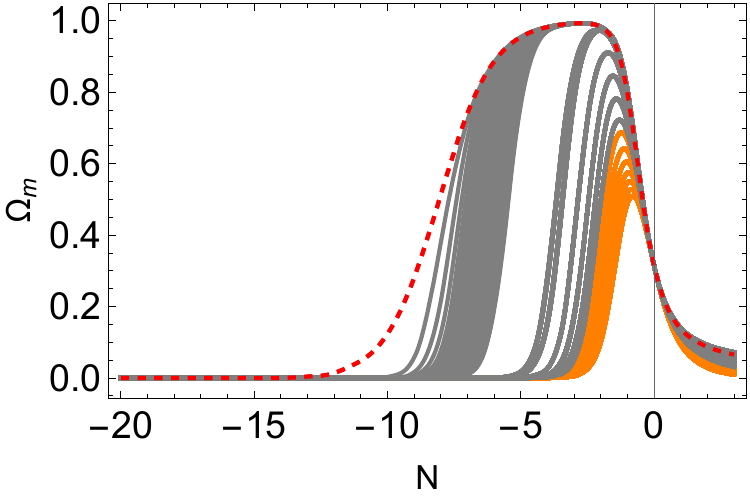} \hspace{0.1in}
\includegraphics[scale = 0.38]{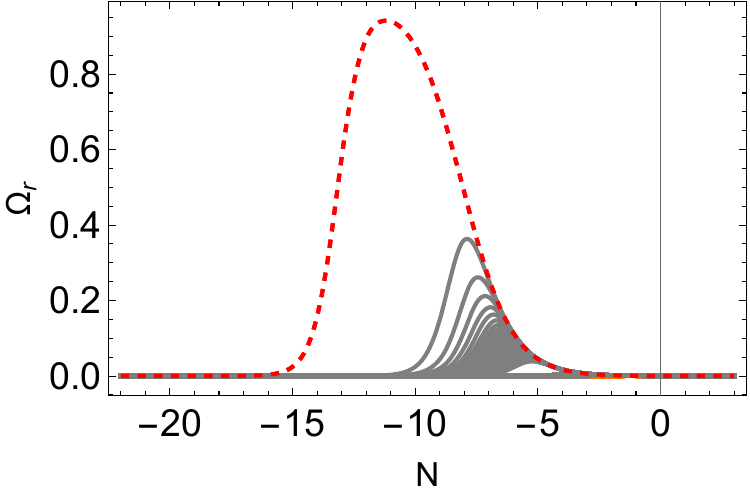} \hspace{0.1in}
\includegraphics[scale = 0.38]{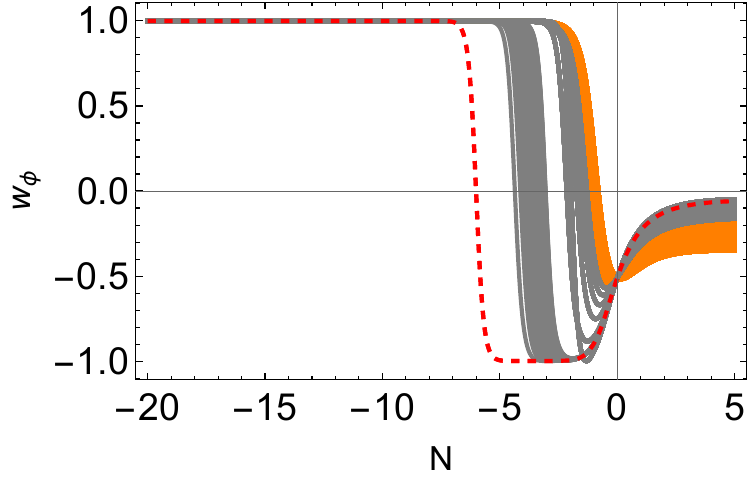} \hspace{0.1in}
\includegraphics[scale = 0.38]{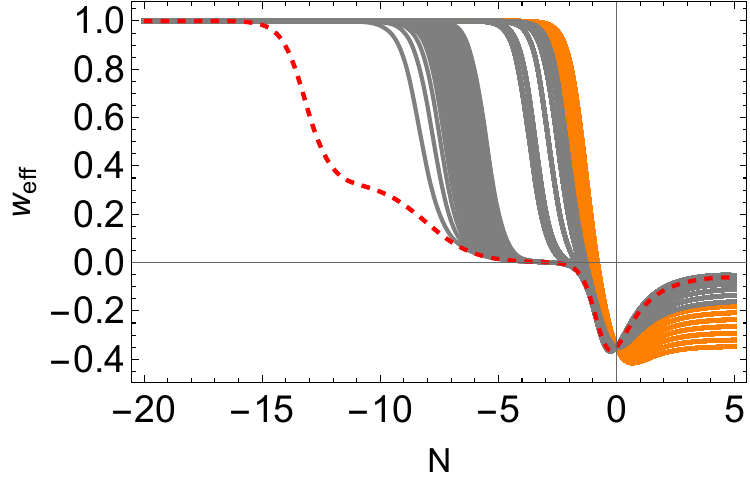}
\caption{The evolution of the cosmological parameters $\Omega_{\phi}$, $\Omega_{k}$, $\Omega_{m}$, and $\Omega_{r}$, together with the scalar-field equation-of-state parameter $w_{\phi}$ and the effective equation-of-state parameter $w_{\mathrm{eff}}$, as a function of the e-folding number $N=ln\,a$. The system is initialized at the present 
epoch with $\Omega_{\phi0} = 0.685$, $\Omega_{r0} = 0.0001$, 
$\Omega_{k0} = 0.0007$. The initial conditions are constructed following the procedure of Ref.~\cite{Andriot:2024jsh}, with the key difference that in the present analysis we vary the slope parameter $\lambda$. 
We explore the parameter range $1.4 < \lambda < 1.6$ (orange curves) and $1.6 < \lambda < 1.75$ (gray curves), consistent with swampland-motivated considerations along with $w_{\phi0} \in [-0.51054453650, -0.510]$. The red dashed curves highlight the representative case $\lambda = \sqrt{3} \approx 1.73205$ and $w_{\phi0} = -0.51054453650$. We shall later disregard the early-universe part of the evolution, e.g. $N<N_{\rm cut}\simeq -11.942$ for $\lambda=\sqrt{3}$, based on the swampland distance conjecture (see Sec.~\ref{sec4}).}
\label{Evolution}
\end{figure*}

The fixed points of the system and their stability properties have been studied in detail in Ref.~\cite{Andriot:2024jsh}. The central motivation for revisiting this framework is to assess whether the model can reproduce a realistic cosmological history, namely a sequence of radiation domination, followed by matter domination, and ultimately a phase of late-time accelerated expansion, as required by observations. Achieving such a sequence becomes increasingly challenging for large values of the slope parameter $\lambda$. In particular, viable trajectories become marginal around $\lambda \simeq 2$, while they are more readily obtained for smaller values of $\lambda$. In other words, imposing a physically reasonable radiation-dominated epoch in the early universe, characterized by $\Omega_r(t) \geq 0.5$, naturally leads to a subsequent period of matter domination whose amplitude and duration are comparable to those in the standard $\Lambda$CDM scenario. 

In the standard spatially flat universe, the scalar-field-dominated attractor yields accelerated expansion only when $\lambda < \sqrt{2}$. Consequently, for $\lambda \geq \sqrt{2}$, the late-time attractor is non-accelerating, although transient acceleration may still arise depending on the initial conditions. Observational analyses of flat exponential quintessence models likewise tend to favor values $\lambda < \sqrt{2}$, making the regime $\lambda \geq \sqrt{2}$ difficult to accommodate within the conventional flat cosmology.  

Motivated by these results, several works have shown that allowing for non-zero spatial curvature qualitatively modifies the phase-space structure. In particular, negatively curved universes ($\Omega_k > 0$) admit cosmologically viable trajectories exhibiting late-time accelerated expansion even for $\lambda > \sqrt{2}$~\cite{Andriot:2024jsh,Agrawal:2018own,Akrami:2018ylq,Raveri:2018ddi,Cline:2001nq}. This makes it possible to explore regions of parameter space that are inaccessible in the spatially flat case. This demonstrates that spatial curvature can restore viable late-time accelerating solutions in a region where the standard flat model fails to do so.

This result has attracted considerable attention, especially in the context of string theory, where exponential potentials arise naturally in low-energy effective descriptions of compactified higher-dimensional theories. In particular, the S2 compactification scenario motivates the parameter range $\lambda \geq \sqrt{3/2}$. Accordingly, in the present work, we perform the observational analysis over the range $\lambda \geq \sqrt{3/2}$ within a non-zero spatial curvature framework. This allows us to test whether curved cosmologies can provide viable late-time accelerating solutions throughout the string-motivated parameter space, which naturally includes $\lambda \geq \sqrt{2}$, where the phase-space structure and late-time cosmological evolution differ qualitatively from those of the standard spatially flat model.

Before confronting the curved quintessence model with cosmological observations, we first solve the autonomous system of differential equations~\eqref{d1}--\eqref{d4} using the present-day initial conditions specified in Fig.~\ref{Evolution}. We then analyze the resulting background evolution for different values of the slope parameter $\lambda$. The figure shows that as $\lambda$ increases, it becomes progressively more difficult for the quintessence model to reproduce a cosmological history consistent with observations. In particular, the effective equation-of-state parameter $w_{\rm eff}$ does not approach sufficiently negative values at late times, thereby failing to match the observationally favored behavior associated with the $\Lambda$CDM model. The numerical evolution indicates that the cosmological viability of steep exponential quintessence is highly sensitive to the slope parameter. Smaller values of $\lambda$ tend to keep the scalar field excessively close to the potential-dominated regime $w_{\phi} \simeq -1$, thereby suppressing the recovery of a sufficiently long radiation-dominated epoch when evolved backward from present-day conditions. On the other hand, larger values of $\lambda$ drive the field too rapidly away from the accelerated regime. Interestingly, with the help of negative spatial curvature ($\Omega_k>0$), trajectories near $\lambda = \sqrt{3}$ provide a nontrivial balance between these competing effects, allowing an extended dark-energy-like phase together with viable radiation- and matter-dominated eras. This behavior motivates a detailed exploration of the parameter region around $\lambda = \sqrt{3}$ in steep swampland-inspired quintessence scenarios.

Moreover, in a system consisting of pure quintessence with negative spatial curvature ($\Omega_k>0$), a new attractor fixed point emerges for $\lambda > \sqrt{2}$, whose associated solution lies at the threshold of accelerated expansion. To illustrate the cosmological evolution in this regime, Fig.~\ref{Evolution} shows the numerical evolution of the density parameters $\Omega_\phi$, $\Omega_k$, $\Omega_m$, and $\Omega_r$, together with the scalar-field equation-of-state parameter $w_\phi$ and the effective equation-of-state parameter $w_{\rm eff}$, as functions of the e-folding number $N$. The solutions are presented for a range of slope parameters $\lambda \in [1,2]$, with the red dashed curve highlighting the representative case $\lambda=\sqrt{3}$.

From Fig.~\ref{Evolution}, it is evident that reproducing a cosmological history consistent with the standard sequence of radiation domination, matter domination, and late-time acceleration (within the regime of field excursion compatible with the swampland distance conjecture, see Sec.~\ref{sec4}) requires increasingly precise tuning of the present-day scalar-field equation-of-state parameter $w_{\phi0}$ as $\lambda$ approaches larger values. In particular, obtaining a sufficiently long radiation-dominated epoch in the past becomes progressively more challenging, and the degree of fine-tuning in $w_{\phi0}$ needed to recover $\Omega_r \gtrsim 0.5$ at early times exceeds what is supported by current observational uncertainties. This illustrates the increasing fine-tuning required for large $\lambda$ in order to reproduce a realistic cosmological evolution. 
We do find solutions that exhibit an extended matter-dominated phase even for large values of $\lambda$. However, these trajectories fail to produce a sufficiently negative effective equation of state at the present epoch, typically yielding $w_{\mathrm{eff}} > -0.5$. As a result, they do not generate the level of late-time acceleration required by observations.

\section{Consistency of the analysis with the swampland distance conjecture} \label{sec4}

%------------------------------------------------------------
% Final paragraph-style write-up WITH vetted citations
% (LaTeX-ready; paste into your draft)
%------------------------------------------------------------

The swampland distance conjecture (SDC) states that when a scalar field traverses a parametrically large geodesic distance in field space, an infinite tower of states becomes exponentially light, indicating the breakdown of a low-energy effective field theory (EFT) description \cite{OoguriVafa2006,Palti2019}. In its standard form the characteristic tower scale behaves as
\begin{equation}
m_{\rm tower}(\phi)\sim m_0\,e^{-\alpha\,\Delta\phi}\,,\qquad \alpha=\mathcal{O}(1),
\label{sdc-tower-final}
\end{equation}
so that sufficiently large $\Delta\phi$ generically necessitates additional light degrees of freedom beyond the single-field EFT. For practical purposes we adopt the conservative requirement that constrains the effective field theory of interest to have a finite number of states
\begin{equation}
m_{\rm tower}(\phi) \gtrsim  \Lambda_{cut} \implies \Delta\phi \lesssim \Delta\phi_{\max}=\alpha^{-1}\ln\Big(\frac{m_0}{\Lambda_{cut}}\Big) \sim \mathcal{O}(1),
\label{sdc-bound-final}
\end{equation}
in reduced Planck units, noting that the precise numerical threshold depends on $\alpha$ and on the cutoff $\Lambda_{cut}$ of the theory given as the threshold of EFT breakdown \cite{Palti2019,OoguriPaltiShiuVafa2019}. 

In a cosmological setting it is natural to implement this criterion in e-fold time $N\equiv \ln a$ by defining an EFT time window $N\in[N_{\rm cut},0]$ and requiring that the scalar excursion accumulated from $N_{\rm cut}$ to today ($N=0$) satisfy Eq.~\eqref{sdc-bound-final}. For its quantitative analysis, we consider the following dimensionless variables defined in Eq.~\eqref{dynamic_sys_variables}
\begin{equation}
x \equiv \frac{\dot\phi}{\sqrt{6}\,H},\qquad \text{and} \qquad
y \equiv \frac{\sqrt{V}}{\sqrt{3}\,H},
\label{xy-def-final}
\end{equation}
so that the scalar density fraction and equation-of-state parameter are
\begin{equation}
\Omega_\phi=x^2+y^2,\qquad 
w_\phi=\frac{x^2-y^2}{x^2+y^2}.
\label{omega-w-final}
\end{equation}
Since $d\phi/dN=\dot\phi/H=\sqrt{6}\,x$, the total excursion between $N_i$ and $N_f$ is exactly
\begin{equation}
\Delta\phi(N_i\!\to\!N_f)
=\int_{N_i}^{N_f}\left|\frac{d\phi}{dN}\right|\,dN
=\sqrt{6}\int_{N_i}^{N_f}|x(N)|\,dN.
\label{DeltaPhi-x-final}
\end{equation}
Moreover, Eq.~\eqref{omega-w-final} implies $(1+w_\phi)\Omega_\phi=(x^2+y^2)+(x^2-y^2)=2x^2$, i.e.\ $x^2=\tfrac{1+w_\phi}{2}\Omega_\phi$, which yields the identity
\begin{equation}
\left|\frac{d\phi}{dN}\right|
=\sqrt{6}\,|x|
=\sqrt{3(1+w_\phi)\Omega_\phi}.
\label{dphidN-omega-w-final}
\end{equation}
Consequently, the excursion can be written purely in terms of $(\Omega_\phi,w_\phi)$ as
\begin{equation}
\Delta\phi(N_i\!\to\!N_f)
=\int_{N_i}^{N_f}\sqrt{3\,(1+w_\phi(N))\,\Omega_\phi(N)}\;dN,
\label{DeltaPhi-omega-w-final}
\end{equation}
so that SDC consistency on the interval $[N_{\rm cut},0]$ reduces to the single integral inequality
\begin{equation}
\Delta\phi(N_{cut}\!\to\!0)
=\int_{N_{\rm cut}}^{0}\sqrt{3\,(1+w_\phi(N))\,\Omega_\phi(N)}\,dN
\le \Delta\phi_{\max}\sim\mathcal{O}(1).
\label{Ncut-criterion-final}
\end{equation}
Equation~\eqref{Ncut-criterion-final} thus provides a direct diagnostic of SDC compatibility using the background evolution displayed in Fig.~\ref{Evolution}. We also note that Ref.~\cite{Andriot:2024sif} established, under mild assumptions, a model-independent upper bound $\Delta\phi\le1$ (in reduced Planck units) for the scalar-field excursion during the late matter--dark energy dominated phase, complementing the criterion presented here. Our criterion is consistent with this result while providing a direct diagnostic based solely on the reconstructed background evolution through $(\Omega_\phi,w_\phi)$.

\begin{figure}[t]
    \centering
    \includegraphics[width=0.7\linewidth]{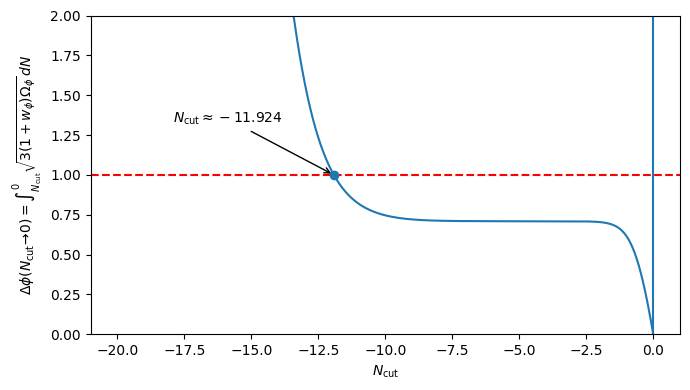}
    \caption{SDC compatibility. The initial conditions and parameters used here are the same as those of Fig. 1.}
    \label{fig_delta_phi_N_cut}
\end{figure}

A central implication is that sufficiently early-time kinetic evolution is generically SDC-unfriendly. Indeed, the dynamical-systems analysis of exponential potentials admits a kinetic-dominated fixed point with $w_\phi\simeq +1$ and $\Omega_\phi\simeq 1$ in the region $N < -15$ as shown in Fig.~\ref{Evolution}. In this limit, Eq.~\eqref{dphidN-omega-w-final} gives an $\mathcal{O}(1)$ displacement per e-fold,
\begin{equation}
\left|\frac{d\phi}{dN}\right|
\simeq \sqrt{3(1+1)\cdot 1}
=\sqrt{6},
\label{dphidN-kinetic-final}
\end{equation}
and hence $\Delta\phi\simeq \sqrt{6}\,\Delta N$ over a duration $\Delta N$ spent in the kinetic regime. Extending the same single-field EFT description deep into such a phase therefore tends to accumulate super-Planckian excursion rapidly and can violate Eq.~\eqref{sdc-bound-final}, signaling the necessity of taking into account additional light degrees of freedom associated with the SDC tower \cite{OoguriVafa2006,Palti2019}. This motivates introducing an EFT {\it initial time} $N_{\rm cut}$ after the end of the kinetic regime, such that the remaining excursion from $N_{\rm cut}$ to $N=0$ is controlled.

For the representative trajectory shown in Fig.~\ref{Evolution} (the $\lambda=\sqrt{3}$ case shown as red dashed curves), the evolution between $N\simeq -11.924$ and $N=0$ contains extended periods where either $w_\phi\approx -1$ (so $1+w_\phi\ll 1$) and/or $\Omega_\phi\ll 1$, suppressing the integrand $\sqrt{3(1+w_\phi)\Omega_\phi}$ in Eq.~\eqref{DeltaPhi-omega-w-final} and hence the accumulated excursion. As a result, shown in Fig.~\ref{fig_delta_phi_N_cut}, the total displacement over $N\in[-11.924,0]$ is expected to be of order unity or smaller and thus plausibly consistent with SDC; this expectation can be verified directly by numerically integrating Eq.~\eqref{DeltaPhi-omega-w-final} once $\Omega_\phi(N)$ and $w_\phi(N)$ (or equivalently $x(N)$) are available. By contrast, pushing further into the much earlier kinetic-dominated regime leads to rapid growth of $\Delta\phi$ according to Eq.~\eqref{dphidN-kinetic-final} and generically worsens SDC compatibility.

According to the SDC, the model introduced in Sec.~\ref{sec3} is not applicable to the earlier epoch. Therefore, in constraining the cosmological parameters by observational data, we restrict our considerations to the later epoch where the model can be consistent with the SDC.

\section{Observational data and methodology}
\label{Data}

In this section, we briefly outline the cosmological datasets and the statistical methodology used to constrain the model discussed above. Specifically, we take into account the following:

\begin{itemize}
    \item CMB:  We make use of temperature and polarization anisotropy measurements of the CMB power spectra from the Planck satellite, in conjunction with their cross-spectra from the 2018 legacy data release of Planck. In particular, we apply the high-$\ell$ Plik likelihood for TT, TE, and EE, along with the low-l likelihood for TT only and the low-$\ell$ EE-only SimAll likelihood. We also include the reconstructed lensing potential obtained from the 3-point correlation function of the Planck data \cite{Planck:2018vyg, Planck:2019nip}.     
    \item DESI BAO: We consider the latest DESI BAO DR2 dataset, which includes observations of baryonic acoustic oscillations found in the clustering patterns of galaxies, quasars, and the Lyman-$\alpha$ forest at high redshifts, grouped into six distinct types of tracers \cite{DESI:2025fii, DESI:2025qqy, DESI:2025zgx}.  
    \item SNeIa Data: Type Ia supernovae (SNeIa) are widely used as standard candles due to their relatively uniform intrinsic luminosity. This data set provides measurements of the apparent magnitude \( m_{b}(z) \), from which the luminosity distance \( D_{L}(z) \) is inferred via the magnitude–redshift relation

\begin{equation}
			\mu \equiv m-M = 5\log(D_L/\text{Mpc}) + 25 \  ,
		\end{equation}
		where, \(m\) denotes the apparent magnitude of the supernova and \(D_L\) is the luminosity distance: 	
		\begin{equation}
			D_L({z}) = c(1+{z}) \int_0^{{z}} \frac{dz'}{H(z')} \ ,
		\end{equation} 
assuming a flat FLRW metric, and \(c\) is the speed of light in km/s. The model parameters are constrained by minimizing the chi-square ($\chi^2$) likelihood, defined as:
		\begin{equation}
			-2 \ln (\mathcal{L}) = \chi^2 = \Delta D^{T} \mathcal{C}^{-1} \Delta D_j\ ,
		\end{equation}
		where $\Delta D = \mu_{\rm Obs} - \mu_{\rm Model}$, $C^{-1}$ denotes the inverse combined statistical and systematic covariance matrix of the SNe sample. We use two different SNe datasets, including PantheonPlus \cite{Brout_2022} and Union3 \cite{DES:2024jxu,DES:2024hip}.
        \begin{itemize}    
    % \item DESY5 Data: This data set consists of Type Ia supernovae observations from the Dark Energy Survey five-year sample (DES-SN5YR), comprising 1829 distinct SNe. It includes 194 nearby SNe with redshift \( z < 0.1 \) and 1635 DES SNe. For our analysis, we compute the likelihood using the distance modulus \( \mu \) and the full covariance matrix provided in the data release.
    \item PP Data: This data set refers to the Pantheon+ compilation, which includes 1550 spectroscopically confirmed Type Ia supernovae. The catalog provides 1770 data samples, from which we use the observational column corresponding to the non-SH0ES-calibrated apparent magnitude \( m_{\rm obs} \). We denote this subset as ``PP'' throughout our analysis.
    \item Union3 Data: It contains more than $1400$ spectroscopically confirmed SNe Ia spanning the redshift range $z \simeq 0.01$--$2.26$, with improved photometric calibration and refined systematic treatments compared to earlier Union releases \cite{Rubin:2023jdq}. The dataset combines observations from several surveys, including SNLS, SDSS, Pan-STARRS, CSP, and multiple low-redshift programs, together with \textit{Hubble Space Telescope} (HST) measurements. 
\end{itemize}

\end{itemize}

We adopt uniform priors for all cosmological parameters. The posterior distributions are sampled using a Markov Chain Monte Carlo (MCMC) approach implemented within the \texttt{COBAYA}\footnote{\href{https://github.com/CobayaSampler/cobaya}{https://github.com/CobayaSampler/cobaya}} framework \cite{Torrado:2020dgo,Torrado:2019cobayaASCL}. The sampling is performed until convergence is achieved according to the Gelman–Rubin criterion \cite{Gelman:1992zz, Brooks:1998}, requiring $R-1 < 0.01$. The resulting chains are subsequently analyzed and visualized using the \texttt{GetDist} package~\cite{Lewis:2019xzd}. The prior ranges for the cosmological parameters employed in the analysis are summarized in Table~\ref{tab:priors} \footnote{We had to choose initial conditions for the field: $\phi=0$ and $\dot{\phi}=0$ deep in the radiation era. A constant shift of the scalar field can always be absorbed into a redefinition of the potential normalization $V_0$, so the choice $\phi_i=0$ is made without loss of generality. Moreover, any finite initial field velocity is rapidly damped by Hubble friction during the radiation-dominated epoch, so the subsequent cosmological evolution is largely insensitive to this choice. For more details, check Ref.~\cite{Schoneberg:2023lun}}. In particular, the priors adopted for the parameter $\lambda$ are motivated by the Swampland $S2$ conjecture as listed in Table~\ref{tab:s1_scenarios_eval_cols}. We note that the $S2$ condition already encapsulates several of the theoretical constraints listed separately, and thus effectively restricts the allowed parameter space in a correlated manner.

\begin{table}[t]
\centering
\begin{tabular}{cc}
\hline\hline
Parameter & Prior range \\
\hline
$\Omega_{k}$ & $[-0.05,0.05]$ \\
$\Omega_{b}h^2$ & $[0.005,\,0.1]$ \\
$\Omega_{c}h^2$ & $[0.001,\,0.99]$ \\
$\lambda$ & $[1.225,2]$ \\
$H_{0}$ & $[50,\,100]$ \\
\hline\hline
\end{tabular}
\caption{Ranges of model parameters employed in our model. The $\bf{S2}$ corresponds to $s_1 = \sqrt{3/2} \approx 1.225$ for the prior of $\lambda$.}
\label{tab:priors}
\end{table}

\begin{figure}[t]
\centering
\includegraphics[scale=0.4]{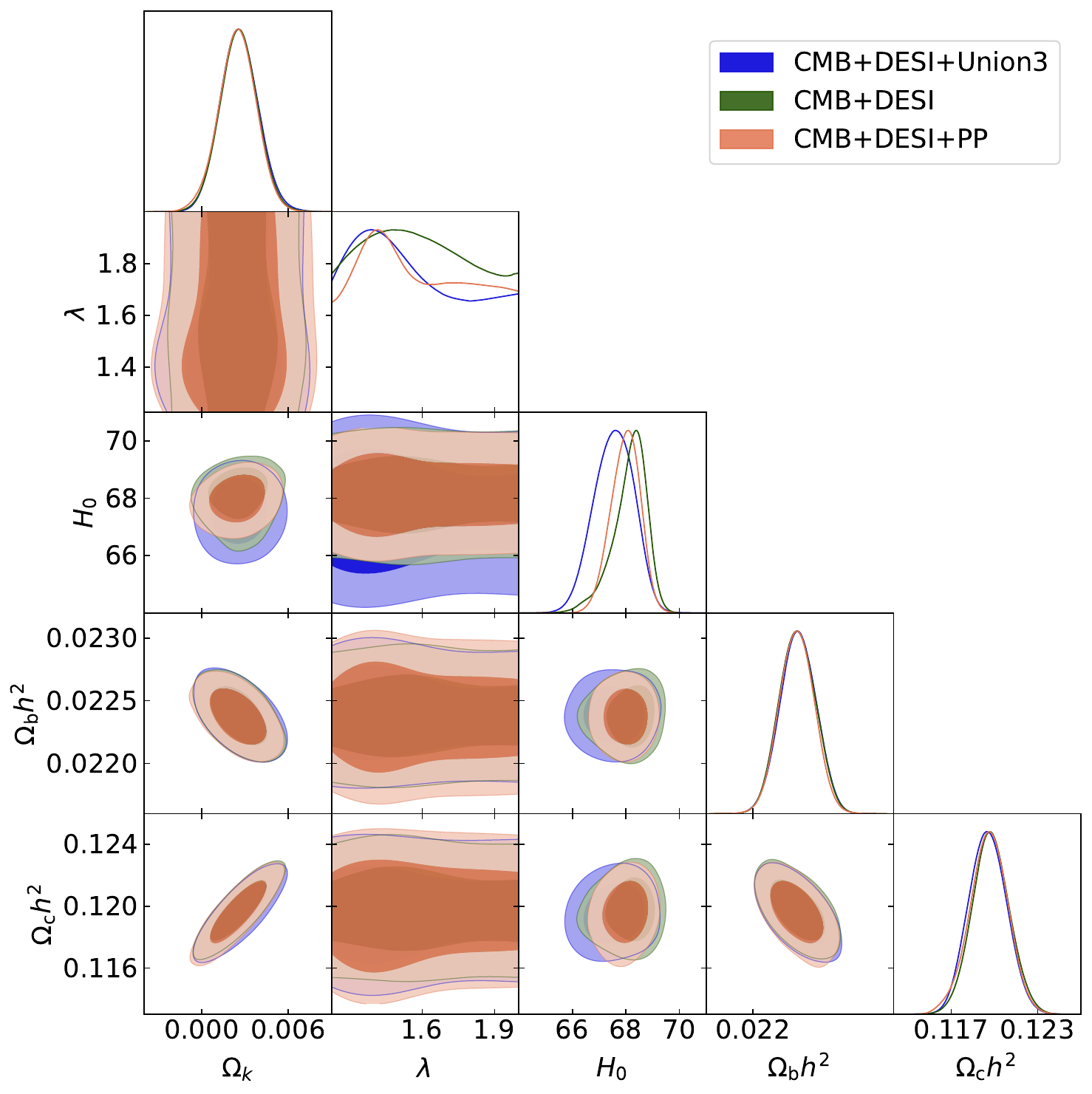}
\caption{Constraints on the set of cosmological parameters for \text{CMB+DESI} along with \text{PP} and \text{Union3}.}
\label{fig:C1}
\end{figure}

The marginalized one- and two-dimensional posterior distributions of the model parameters are shown in Fig.~\ref{fig:C1}. The parameter $\lambda$ is treated as a free parameter of the quintessence model, while the potential normalization $V_0$ is getting tuned at each of the MCMC to satisfy the budget Eq~\eqref{Omega_m_constraint}. We perform the analysis using the dataset combinations $CMB+DESI$, $CMB+DESI+Pantheon+$ and $CMB+DESI+Union3$ for the curved (i.e, nonzero $\Omega_k$). As reported in Table~\ref{tab:cosmo_params}, the current data do not place significant constraints on the quintessence parameter $\lambda$. Nevertheless, a mild preference for positive curvature $\Omega_{k}>0$ is observed across all dataset combinations.

The standard cosmological parameters are well constrained and remain consistent with the values inferred within the $\Lambda$CDM framework. In particular, the constraints on the other cosmological parameters closely follow those obtained in previous analyses using similar datasets. On the other hand, the parameter $\lambda$, which characterizes the slope of the exponential quintessence potential, remains weakly constrained by the data and spans a large fraction of the allowed prior range. Consequently, cosmological observables primarily constrain the overall background evolution while leaving the slope of the potential largely degenerate. As a result, the posterior distribution of $\lambda$ remains partially prior-dominated, indicating that current datasets lack the sensitivity required to tightly constrain the steepness of the potential within the theoretically motivated swampland prior range.

These findings are consistent with recent studies of similar scenarios~\cite{Alestas:2024gxe}. In particular, although curvature-assisted quintessence models allow for slightly larger values of $\lambda$ compared to the spatially flat case, the corresponding enlargement of the viable parameter space is modest. Nevertheless, the primary objective of the present work is not to establish full compatibility of the model with Swampland expectations, but rather to investigate whether theoretically-motivated priors can lead to observable shifts in cosmological parameter inference relative to the standard analysis employing broad, theory-agnostic priors. From this perspective, our results indicate that the Swampland-inspired restrictions can indeed mildly affect the inferred posterior distribution of $\Omega_k$, while leaving the overall observational viability of the model broadly consistent with current data.

\begin{table}[t]
\centering
\begin{tabular}{lccc}
\toprule
\textbf{Parameter} & \textbf{CMB+DESI} & \textbf{CMB+DESI+PP} & \textbf{CMB+DESI+Union3} \\
\midrule
$\Omega_b\,h^2$ & $0.02238 \pm 0.00015$ & $0.02237 \pm 0.00015$ & $0.02238 \pm 0.00015$ \\ [0.2em]
$\Omega_c\,h^2$ & $0.1197 \pm 0.0013$ & $0.1196 \pm 0.0013$ & $0.1195 \pm 0.0013$ \\ [0.2em]
$ln(10^{10}A_s)$ & $3.053 \pm 0.015$ & $3.055^{+0.014}_{-0.015}$ & $3.054\pm 0.015$ \\ [0.2em]
$n_s$ & $0.9652 \pm 0.0043$ & $0.9651^{+0.0041}_{-0.0047}$ & $0.9656 \pm 0.0044$ \\ [0.2em]
$\tau_{reio}$ & $0.0588^{+0.0071}_{-0.0079}$ & $0.0599^{+0.0070}_{-0.0096}$ & $0.0590 \pm 0.0076$ \\ [0.2em]
$H_0\;(\mathrm{km\,s^{-1}Mpc^{-1}})$ & $68.14^{+0.74}_{-0.42}$ & $67.98^{+0.59}_{-0.50}$ & $67.55^{+0.81}_{-0.69}$ \\ [0.2em]
$\lambda$ & $ < 1.60$ & $1.60^{+0.23}_{-0.31}$ & $ < 1.58$ \\ [0.2em]
$\Omega_k$ & $0.0025 \pm 0.0013$ & $0.0025 \pm 0.0013$ & $0.0026 \pm 0.0013$ \\ [0.2em]
$r_d$ & $147.17 \pm 0.28$ & $147.20 \pm 0.30$ & $147.20 \pm 0.28$ \\
\bottomrule
\end{tabular}
\caption{Observational constraints at 68\% CL on key cosmological parameters obtained from different datasets.}
\label{tab:cosmo_params}
\end{table}

We now compare the performance of the curved quintessence model with that of the non-flat $\Lambda$CDM model ($\Lambda$CDM + $\Omega_k$) against observational data. The best-fit parameters are obtained using the \textsc{minimize} sampler by minimizing the chi-square, $\chi^2$. To quantify the relative goodness of fit, we compute the difference
\begin{equation}
\Delta \chi^2_{\text{model}} = \chi^2_{\text{model}} - \chi^2_{\Lambda \mathrm{CDM}}\,,
\end{equation}
where $\chi^2 = -2 \ln \mathcal{L}_{\text{max}}$ for a given model. A negative value of $\Delta \chi^2$ indicates a better fit relative to $\Lambda$CDM, while a positive value indicates a worse fit.

We obtain $\Delta \chi^2 = -0.23, -2.79, -4.18$ for the dataset combinations \text{CMB+DESI}, \text{CMB+DESI+PP}, and \text{CMB+DESI+Union3}, respectively. While these values formally indicate an improved fit compared to $\Lambda$CDM, the improvement is modest and should be interpreted with caution. In particular, the quintessence model contains additional freedom and can recover the $\Lambda$CDM limit in the regime $\lambda \to 0$, which partially explains the reduction in $\chi^2$.

To account for differences in model complexity, we also evaluate the Akaike Information Criterion (AIC)~\cite{Akaike:1974vps,Schwarz:1978tpv}, defined as $\mathrm{AIC} = -2 \ln \mathcal{L}_{\text{max}} + 2k$, where $k$ denotes the number of free parameters. We find $\Delta \mathrm{AIC} = 1.77, 0.79, -2.18$ for the same dataset combinations. These values indicate that, after penalizing for the additional model complexity, there is no strong statistical preference for the curved quintessence model over $\Lambda$CDM. At most, the $\mathrm{CMB}+\mathrm{DESI}+\mathrm{Union3}$ combination shows a mild preference, but this remains statistically inconclusive. 

Our inferred value of the curvature parameter may also be compared with the recent DESI analysis of the $w_0w_a+\Omega_k$ parametrization~\cite{DESI:2025qqy}. Although the CPL parametrization and exponential quintessence represent different descriptions of dark energy, both analyses favor a small positive value of $\Omega_k$. For instance, the DESI+CMB+PP analysis reports $\Omega_k = 0.0011 \pm 0.0013$ in the recent DESI DR2, while the present analysis yields $\Omega_k = 0.0025 \pm 0.0013$. A similarly good agreement is obtained for the DESI+CMB+Union3 dataset combination, although some DESI dataset combinations instead prefer a mildly closed universe. Overall, our constraints on the spatial curvature are consistent with those inferred from the DESI analysis despite the different underlying dark-energy models. This behavior is also qualitatively consistent with earlier studies of exponential quintessence in non-flat cosmologies, which found that allowing for non-zero spatial curvature enlarges the viable parameter space of exponential potentials~\cite{Andriot:2024jsh,Bhattacharya:2024hep}.

As previously mentioned, our primary interest in the present paper is the possibility that theoretically motivated priors may shift values of cosmological parameters inferred by observational data, compared with the standard analysis based on theory-agnostic priors such as a sufficiently wide flat prior. Indeed, the (curved) $\Lambda$CDM limit $\lambda \to 0$ is excluded by the swampland-motivated prior and is therefore disconnected from the model considered in the present work. If such theoretical priors are taken seriously, a natural question is whether the ranges of cosmological parameters supported by observational data, particularly $\Omega_k$, differ from those obtained in the conventional (curved) $\Lambda$CDM framework. We find that the swampland-motivated prior leads to a mild shift in the preferred posterior region of $\Omega_k$, although the effect is not statistically significant with current data. Nevertheless, this illustrates that theoretically-informed priors can leave an observable imprint on cosmological parameter inference and therefore deserve careful consideration in phenomenological analyses of dark energy models.

\section{Conclusion} \label{conc}

In this work, we have examined whether allowing for nonzero spatial curvature can render single-field quintessence models with steep exponential potentials observationally viable. This question is well motivated by recent studies suggesting that curvature effects may alter the late-time attractor structure and, in principle, support accelerated expansion in regimes that are otherwise disfavored in spatially flat cosmologies. Motivated by these considerations, we formulated the dynamical system describing exponential quintessence in a non-flat universe and investigated its cosmological evolution numerically. In parallel, we explored the observational implications of imposing Swampland-motivated theoretical priors on the slope parameter $\lambda$, which exclude the (curved) $\Lambda$CDM limit $\lambda \to 0$ and therefore lead to a phenomenologically distinct framework compared to standard dark energy analyses employing broad theory-agnostic priors. Our primary aim was to investigate whether such theoretically-informed priors can induce observable shifts in cosmological parameter inference, particularly in the preferred range of the spatial curvature parameter $\Omega_k$.

As discussed in Sec.~\ref{conjectures}, we incorporate theoretical consistency conditions motivated by the de Sitter conjecture and asymptotic trans-Planckian censorship bounds by introducing a set of order-one parameters that quantify the strength of the UV-consistency conditions. Given the theoretical uncertainty associated with these parameters, we adopt a scenario-based approach in which a representative set of theoretically motivated values for $s_1$ (summarized in Table~\ref{tab:s1_scenarios}) are treated as alternative theory priors. This framework allows us to systematically assess the robustness of our cosmological results, particularly those concerning the preferred sign and magnitude of spatial curvature, under different realizations of the swampland constraints.

We have focused on a single-field quintessence model of dark energy with an exponential potential and explored its phase-space behavior numerically. Our results show that, although the inclusion of spatial curvature modestly enlarges the allowed range of the slope parameter $\lambda$ compared to the spatially flat case, this extension is insufficient to reconcile the model simultaneously with observational constraints and the theoretically motivated lower bounds on $\lambda$. In particular, current cosmological observations favor the lower end of the string-motivated parameter range, while progressively larger values of $\lambda$ provide a less favorable fit to the data. Nevertheless, the overall statistical performance of the model remains comparable to that of $\Lambda$CDM for the dataset combinations considered.

% In particular, regions of parameter space accommodating larger values of $\lambda$ fail to provide an adequate fit to current cosmological data, pointing to a persistent tension between steep-potential quintessence and observational viability.

The dynamical evolution further clarifies this behaviour. While spatial curvature alters the background evolution and can shift the onset of scalar-field domination, our analysis indicates that the region of parameter space associated with the larger values of $\lambda$ allowed in this work exhibits a progressively shorter matter-dominated epoch. Consequently, the late-time effective equation of state departs from values close to $w_{\rm eff} \simeq -1$, remaining around $w_{\rm eff} \gtrsim -0.5$. These features contribute to the comparatively poorer agreement with current cosmological observations found in this region of the parameter space.

% The dynamical evolution further clarifies this behavior. While spatial curvature alters the background evolution and can shift the onset of scalar-field domination, the resulting cosmological history does not sustain a sufficiently prolonged matter-dominated epoch for larger values of $\lambda$. The duration of the matter era is reduced, with potential implications for structure formation, and the effective equation-of-state parameter does not approach values close to $w_{\rm eff} \simeq -1$, instead remaining around $w_{\rm eff} \gtrsim -0.7$ in these regimes. These features provide a dynamical explanation for why curvature-assisted solutions do not translate into observationally viable scenarios for steep exponential quintessence.

As shown in Fig.~\ref{fig:C1}, the parameter $\lambda$ remains only weakly constrained across all dataset combinations, with the posterior spanning a large fraction of the allowed prior range. This indicates that current cosmological observations lack the sensitivity required to tightly constrain the steepness of the exponential potential within this framework. We also find that the constraints on the spatial curvature parameter $\Omega_k$ exhibit a mild preference for an open universe, although this trend is not statistically significant. Taken together, these results reinforce the conclusion that the inclusion of spatial curvature does not substantially enhance the observational viability of steep-potential quintessence models. Instead, the allowed parameter space remains largely driven by the imposed theoretical priors rather than by the data. In particular, we do not find evidence that allowing for nonzero spatial curvature leads to a statistically meaningful improvement in the fit to current observations. This suggests that more general scalar-field constructions, additional degrees of freedom, or extended dark-sector interactions may be required to achieve a viable description of cosmic acceleration consistent with UV-motivated expectations. Future high-precision cosmological observations will play a crucial role in further testing these scenarios and clarifying the extent to which curvature or more complex dynamics can alleviate the tension between theory and data.

\subsection*{Acknowledgments} 
SA acknowledges the Japan Society for the Promotion of Science (JSPS) for providing a postdoctoral fellowship during 2024-2026 (JSPS ID No.: P24318). This work of SA is supported by the JSPS KAKENHI grant (Number: 24KF0229). The work of SM was supported in part by JSPS KAKENHI Grant No.\ JP24K07017 and World Premier International Research Center Initiative (WPI), MEXT, Japan.
HJ is supported by the National Research Foundation of Korea (NRF) through the Grants: RS-2020-NR049598 (CQUeST, Sogang University), RS-2023-NR077094, and RS-2024-00441954.

\appendix

% ============================================================
% Derivations (English, LaTeX) for Eqs. (I.1) and related
% distance relations: D_M = (1+z) D_A and D_L = (1+z) D_M
% ============================================================

\section{Derivation of the comoving transverse distance in curved FLRW}\label{appendix}

Consider the FLRW line element in standard curvature coordinates
\begin{equation}
ds^2 = -c^2 dt^2
+ a^2(t)\left[\frac{dr^2}{1-k r^2} + r^2 d\Omega^2 \right],
\qquad k\in\{+1,0,-1\},
\label{flrw_r}
\end{equation}
where we have introduced the shorthand
\begin{equation}
d\Omega^2 \equiv d\theta^2 + \sin^2\theta\, d\phi^2.
\end{equation}

For a radial light ray, we have $ds^2 = 0$ and $d\Omega = 0$. Equation~\eqref{flrw_r} then reduces to
\begin{equation}
0 = -c^2 dt^2 + a^2(t)\frac{dr^2}{1-k r^2}
\quad\Longrightarrow\quad
c^2 dt^2 = a^2(t)\frac{dr^2}{1-k r^2}.
\end{equation}
Taking the square root, we obtain
\begin{equation}
c\,dt = \pm a(t)\frac{dr}{\sqrt{1-k r^2}},
\label{null_root}
\end{equation}
where the sign corresponds to ingoing or outgoing null geodesics.\\

We now define a new radial coordinate $\chi$ via
\begin{equation}
d\chi \equiv \frac{dr}{\sqrt{1-k r^2}},
\label{chi_def}
\end{equation}
so that Eq.~\eqref{null_root} becomes
\begin{equation}
d\chi = \pm \frac{c\,dt}{a(t)}.
\label{dchi_dt}
\end{equation}

Integrating Eq.~\eqref{chi_def} with the boundary condition $r=0 \leftrightarrow \chi=0$, we obtain
\begin{equation}
r = S_k(\chi), \qquad
S_k(\chi)=
\begin{cases}
\sin\chi, & k=+1,\\[4pt]
\chi, & k=0,\\[4pt]
\sinh\chi, & k=-1.
\end{cases}
\label{Sk_def}
\end{equation}

With \eqref{chi_def}--\eqref{Sk_def}, the metric can be written as
\begin{equation}
ds^2 = -c^2 dt^2 + a^2(t)\Big[d\chi^2 + S_k^2(\chi)\,d\Omega^2\Big].
\label{flrw_chi}
\end{equation}

$\chi(z)$ as a redshift integral. Choose the sign and limits so that $\chi(z)\ge 0$ denotes the comoving radial separation
between an emitter at time $t_e$ (redshift $z$) and the observer at $t_0$ (redshift $0$):
\begin{equation}
\chi(z) \equiv \int_{t_e}^{t_0}\frac{c\,dt}{a(t)}.
\label{chi_integral_t}
\end{equation}

To express the null geodesic relation in terms of observable quantities, it is convenient to rewrite it in terms of redshift. The redshift is defined as
\begin{equation}
1+z \equiv \frac{a_0}{a(t)}, \qquad a_0 \equiv a(t_0),
\label{redshift_def}
\end{equation}
where $t_0$ denotes the present time. The Hubble parameter is given by $H(t)\equiv \dot a/a$. Differentiating Eq.~\eqref{redshift_def}, we obtain
\begin{equation}
\frac{dz}{dt} = -(1+z)H(t),
\end{equation}
which can be rearranged as
\begin{equation}
dt = -\frac{dz}{(1+z)H(z)}.
\label{dt_dz}
\end{equation}

Using $a(t)=a_0/(1+z)$, we can rewrite the integrand appearing in the null geodesic relation as
\begin{equation}
\frac{dt}{a(t)} = -\frac{dz}{a_0 H(z)}.
\label{dt_over_a}
\end{equation}

Substituting this into Eq.~\eqref{chi_integral_t} and noting that $t$ increases from $t_e$ to $t_0$ while $z$ decreases from $z$ to $0$, we obtain
\begin{equation}
\chi(z) = \int_{t_e}^{t_0} \frac{c\,dt}{a(t)} 
= \frac{c}{a_0} \int_{0}^{z} \frac{dz'}{H(z')}.
\label{chi_of_z}
\end{equation}

\noindent {\bf Angular-diameter distance and comoving transverse distance.}
To relate the metric to observable distances, consider transverse separations at fixed cosmic time and radial coordinate. From Eq.~\eqref{flrw_chi}, setting $dt=0$ and $d\chi=0$ (and, for simplicity, $d\phi=0$), we obtain
\begin{equation}
ds^2 = a^2(t) S_k^2(\chi)\, d\theta^2
\quad\Longrightarrow\quad
d\ell_\perp = a(t) S_k(\chi)\, d\theta.
\label{dl_perp}
\end{equation}
The angular-diameter distance is defined as
\begin{equation}
D_A \equiv \frac{d\ell_\perp}{d\theta},
\end{equation}
which gives
\begin{equation}
D_A(t) = a(t)\, S_k(\chi).
\label{DA_def}
\end{equation}

For a source at redshift $z$ (emitted at time $t_e$), this becomes
\begin{equation}
D_A(z) = a(t_e)\, S_k\bigl(\chi(z)\bigr).
\label{DA_z}
\end{equation}

It is often convenient to introduce the comoving transverse distance, defined as
\begin{equation}
D_M(z) \equiv (1+z)\, D_A(z).
\label{DM_def}
\end{equation}
Using $a(t_e)=a_0/(1+z)$ from Eq.~\eqref{redshift_def}, we obtain
\begin{equation}
D_A(z) = \frac{a_0}{1+z} S_k\bigl(\chi(z)\bigr),
\quad\Longrightarrow\quad
D_M(z) = a_0\, S_k\bigl(\chi(z)\bigr).
\label{DM_final}
\end{equation}

Substituting Eq.~\eqref{chi_of_z} into Eq.~\eqref{DM_final}, we obtain
\begin{equation}
D_M(z) = a_0\, S_k\!\left(\frac{c}{a_0}\int_{0}^{z}\frac{dz'}{H(z')}\right).
\label{DM_Hint}
\end{equation}

\noindent {\bf Rewriting in terms of $\Omega_k$ and $E(z)$.}
We introduce the curvature density parameter at the present epoch,
\begin{equation}
\Omega_k \equiv -\frac{k\,c^2}{a_0^2 H_0^2},
\qquad H_0 \equiv H(z{=}0),
\label{Omegak_def}
\end{equation}
and define the dimensionless Hubble rate
\begin{equation}
E(z)\equiv \frac{H(z)}{H_0},
\quad\Longrightarrow\quad
\int_0^z\frac{dz'}{H(z')}=\frac{1}{H_0}\int_0^z\frac{dz'}{E(z')}.
\label{E_def}
\end{equation}

For $k=\pm1$, Eq.~\eqref{Omegak_def} implies
\begin{equation}
\sqrt{|\Omega_k|}=\frac{c}{a_0 H_0}
\quad\Longleftrightarrow\quad
a_0=\frac{c}{H_0\sqrt{|\Omega_k|}}.
\label{curv_radius_rel}
\end{equation}

Substituting Eqs.~\eqref{E_def} and \eqref{curv_radius_rel} into Eq.~\eqref{DM_Hint}, the argument of $S_k$ becomes
\begin{equation}
\frac{c}{a_0}\int_0^z\frac{dz'}{H(z')}
=
\sqrt{|\Omega_k|}\int_0^z\frac{dz'}{E(z')},
\label{arg_rewrite}
\end{equation}
while the prefactor $a_0$ is expressed in terms of $\Omega_k$. We thus obtain
\begin{equation}
D_M(z)
=
\frac{c}{H_0}\frac{1}{\sqrt{|\Omega_k|}}
\,S_k\!\left(\sqrt{|\Omega_k|}\int_0^z\frac{dz'}{E(z')}\right),
\label{DM_final2}
\end{equation}
which is Eq.~(I.1).

\paragraph{Flat limit.}
As $\Omega_k\to 0$, $S_k(x)\to x$, and
\[
\frac{1}{\sqrt{|\Omega_k|}}S_k\!\left(\sqrt{|\Omega_k|}A\right)\to A,
\]
so \eqref{DM_final2} reduces to
\[
D_M(z)\to \frac{c}{H_0}\int_0^z\frac{dz'}{E(z')}.
\]

\bibliographystyle{unsrtnat}
\bibliography{Ref}

\end{document}